\documentclass[reqno, 12pt]{amsart}

\usepackage{amsmath,amssymb,braket,appendix,nicefrac,enumitem}
\usepackage[protrusion=true,babel=true]{microtype}
\usepackage[english]{babel}
\usepackage[widespace]{fourier}
\usepackage[backrefs]{amsrefs}
\usepackage[margin=1in]{geometry}
\usepackage[onehalfspacing]{setspace}
\usepackage[pdfusetitle, pagebackref]{hyperref}

\usepackage{graphicx}

\numberwithin{equation}{section}

\usepackage[nameinlink, noabbrev]{cleveref}
\expandafter\def\csname ver@etex.sty\endcsname{3000/12/31}

\usepackage{autonum}

\allowdisplaybreaks[1]

\let\originalleft\left
\let\originalright\right
\renewcommand{\left}{\mathopen{}\mathclose\bgroup\originalleft}
\renewcommand{\right}{\aftergroup\egroup\originalright}

\def\({\mathopen{}\left(}
\def\){\right)\mathclose{}}

\makeatletter
\renewcommand*{\eqref}[1]{\hyperref[{#1}]{\textup{\tagform@{\ref*{#1}}}}}
\makeatother

\newcommand*{\eqdef}{\mathrel{\vcenter{\baselineskip0.5ex \lineskiplimit0pt\hbox{.}\hbox{.}}}=}
\newcommand*{\defeq}{=\mathrel{\vcenter{\baselineskip0.5ex \lineskiplimit0pt\hbox{.}\hbox{.}}}}

\newtheorem{theorem}{Theorem}[section]
\newtheorem{remark}[theorem]{Remark}

\crefname{remark}{Remark}{Remarks}
\creflabelformat{remark}{#2{#1}#3}
\crefname{oracle}{Oracle}{Oracles}
\creflabelformat{oracle}{#2{\upshape #1}#3}

\def\F{\mathbb{F}}
\def\G{\mathbb{G}}
\def\rl{\mathbb{R}}
\def\P{\mathbb{P}}
\def\Z{\mathbb{Z}}

\def\A{\mathtt{A}}
\def\CNOT{\mathrm{CNOT}}
\def\cO{\mathcal{O}}
\def\PFO{\mathtt{PFO}}
\def\sinc{\mathrm{sinc}}
\def\supp{\mathrm{supp}}
\def\QFT{\mathrm{QFT}}
\def\W{\mathcal{W}}
\def\WH{\mathrm{wh}}

\title{Novel oracle constructions for quantum random access memory}
\date{\today}
\keywords{quantum memory, quantum dictionaries, data-access oracles, Walsh--Hadamard Transform}

\author{\'Akos Nagy}
\address[\'Akos Nagy]{BEIT Canada, Toronto, Ontario}
\email{\href{mailto:contact@akosnagy.com}{contact@akosnagy.com}}
\urladdr{\href{https://akosnagy.com/}{akosnagy.com}}

\author{Cindy Zhang}
\address[Cindy Zhang]{}
\email{\href{mailto:xindizhang.phys@gmail.com}{xindizhang.phys@gmail.com}}

\hypersetup{
    unicode         = true,
    pdffitwindow    = true,
    pdftoolbar      = false,
    pdfmenubar      = false,
    pdfstartview    = {FitH},
    hypertexnames   = true,
    colorlinks      = true,
    linkcolor       = black,
    citecolor       = black,
    filecolor       = black,
    urlcolor        = blue
}

\calclayout
\begin{document}

\begin{abstract}
    We present new designs for quantum random access memory. More precisely, for each function, $f : \F_2^n \rightarrow \F_2^d$, we construct oracles, $\cO_f$, with the property
    \begin{equation}
        \cO_f \ket{x}_n \ket{0}_d = \ket{x}_n \ket{f(x)}_d.
    \end{equation}
    Our methods are based on the Walsh--Hadamard Transform of $f$, viewed as an integer valued function. In general, the complexity of our method scales with the sparsity of the Walsh--Hadamard Transform and not the sparsity of $f$, yielding more favorable constructions in cases such as binary optimization problems and function with low-degree Walsh--Hadamard Transforms. Furthermore, our design comes with a tuneable amount of ancillas that can trade depth for size. In the ancilla-free design, these oracles can be $\epsilon$-approximated so that the Clifford + $T$ depth is $O \( \( n + \log_2 (\nicefrac{d}{\epsilon}) \) \W_f \)$, where $\W_f$ is the number of nonzero components in the Walsh--Hadamard Transform. The depth of the shallowest version is $O \( n + \log_2 (\nicefrac{d}{\epsilon}) \)$, using $n + d \W_f$ qubit. The connectivity of these circuits is also only logarithmic in $\W_f$. As an application, we show that for boolean functions with low approximate degrees (as in the case of read-once formulas) the complexities of the corresponding QRAM oracles scale only as $2^{\widetilde{O} \( \sqrt{n} \log_2 \( n \) \)}$.
\end{abstract}

\maketitle

\section{Introduction}

The construction of complex quantum oracles is a central problem in quantum computing. Given a binary function, $f : \F_2^n \rightarrow \F_2^d$, an oracle that acts on $n + d$ qubits as
\begin{equation}
    \cO_f \ket{x}_n \ket{0}_d = \ket{x}_n \ket{f(x)}_d, \label{eq:qram_def}
\end{equation}
is called a \emph{quantum random access classical memory (QRAM) oracle}, or more precisely, quantum random access \emph{classical} memory; cf. \cite{jaques_qram_2023,phalak_quantum_2023}. These oracles are also known as quantum dictionary encoders or data-access oracles and are used as subroutines in many quantum algorithms, such as binary optimization problems \cite{gilliam_grover_2021}, state preparation for simulation \cite{lee_even_2021}, or as the marker oracle in Grover's algorithm \cite{park_t-depth_2023}.

\smallskip

Resource-efficient implementations of $\cO_f$ is an active area of research. Despite its simple definition, the decomposition is not trivial, especially when one is restricted to implementing this oracle using only elements of a limited, but universal gate set, say only $1$-qubit $z$-rotations and $\CNOT$ gates. Known implementations are typically based on the decomposition of the function $f$ in certain bases; for example, \cite{low_trading_2018} uses the decomposition in the standard basis, while \cite{seidel_automatic_2023} uses the Walsh--Hadamard Transform of the components of $f$, viewed as a binary vector. In both cases, $\boxplus = \oplus$ is the componentwise addition modulo $2$. As our designs implement $\cO_f$ with $\boxplus = +$ being addition modulo $2^d$, they can yield an advantage in cases where this is preferred. Here we mentioned two such situations. First, our quantum RAM can be seen as a ``controlled-adder'', where $f(x) = y$ means adding $y$ to the last $d$ qubits when the first $n$ are in the state $\ket{x}_n$. The second advantage it yields is that this design favors errors that are numerically close to the correct value over errors that close in Hamming distance. That is, in a noisy implementation on our circuits, if again, $f(x) = y$, then the most likely errors are of the form $\ket{x}_n \ket{y \pm 1}_d$, even if the Hamming distance of $y$ and $y \pm 1$ is large; see more in \Cref{sec:rvf}.

\smallskip

In this paper, we give novel constructions for quantum RAM oracles, based on the Walsh--Hadamard Transform of $f$ (see \cref{eq:WHT}) and the $d$-qubit Quantum Fourier Transform adder
\begin{equation}
    T_{k, d}^\QFT \eqdef \QFT_d^\dagger \circ P(k) \circ \QFT_d, \label{eq:QFT_adder}
\end{equation}
where $P(k)$ is defined so that $P(k) \ket{y}_d = \exp \( \tfrac{2 \pi i}{2^d} ky \) \ket{y}_d$. A key advantage of this approach is that aggregation of the summands happen automatically and instantaneously at the level of phases, thus it is very well-suited for parallelization. In this case, $\boxplus$ is the addition modulo $2^d$. We show that in certain cases (when the Walsh--Hadamard Transform of $f$ is sparse) our design has significantly improved complexities, when compared to previously known methods. For example, in the first oracle design, \cref{oracle:oracle_1} in \Cref{sec:oracles}, the time complexity of one oracle call can be made as short as $O \( n + \log_2 (d) \)$, which remains true even after Clifford + $T$ approximation with an error that is exponentially small in $n$ (see \Cref{sec:fault}). This is an $n$-fold improvement compared to, for example, the best bound achieved in \cite{low_trading_2018}, which is $O \( n^2 + \log_2 (d) \)$ only; see \cite{low_trading_2018}*{Theorems~$1$~\&~$2$}.

\smallskip

Binary functions with ``low-degree'' Walsh--Hadamard Transforms appear in several contexts, such as in deep neural networks. It has been shown that deep networks trained to fit binary functions tend to be biased towards the low-degree components, which connects directly to theories of generalizability; cf. \cites{valle_deep_2018,yang_fine_2019,huh_low_2021}. Our \cref{oracle:oracle_4} is especially well-suited for such functions. Its depth and gate counts scale as $O \( 2^{H_2 \( \nicefrac{k}{n} \) n} \)$, where $H_2$ is the binary entropy (see \cref{eq:entropy}). When $k \ll \tfrac{n}{2}$, this is much smaller than $2^n$, the expected complexities for other methods, such as the SELECT method in \cite{low_trading_2018}.

\smallskip

\subsection*{Organization of the paper:} In \Cref{sec:oracles}, we outline our oracles and their properties and in \Cref{sec:design}, we give the corresponding designs. Finally, in \Cref{sec:fault}, we discuss fault-tolerant implementations and give Clifford + $T$ complexities. In \Cref{sec:approx_degree}, we show how our constructions further simplify for boolean functions with low approximate degrees. In \Cref{sec:WH_adder}, we briefly outline an alternative design with better $T$-complexities in some cases. The designs are verified in \Cref{app:verifications}.

\subsection*{Acknowledgments:} The authors are thankful to Tom Ginsberg, Jan Tu\l{}owiecki, Emil {Ż}{{\. Z}}ak, Szymon Pli\'s, and Raphael Seidel for their valuable feedback. The first author thanks Witold Jarnicki for suggesting the idea for the alternative design of \Cref{sec:WH_adder}.

\subsection*{Supplementary material:} A Github repository containing Qiskit code for the oracles designs can be at \cite{nagy_github_WalshHadamardQRAM}.

\bigskip

\section{Quantum RAM oracles using the Walsh--Hadamard Transform}
\label{sec:oracles}

All oracles in this section are constructed using the gate set $\left\{ H, \CNOT, R_Z, \QFT_d, \QFT_d^\dagger \right\}$, where $R_Z$ denotes single-qubit rotations around the $z$-axis of the Bloch sphere and $\QFT_d$ denotes the $d$-qubit Quantum Fourier Transform. All complexities below are on the top of the Quantum Fourier Transform and its inverse. By \emph{Connectivity} below, we mean the maximal number of qubits each qubit can be directly entangled to via a $\CNOT$ gate.

Design details are included in \Cref{sec:design}, proofs in \Cref{app:verifications}, and fault-tolerant implementation in \Cref{sec:fault}. For the explanation on notations, please see \Cref{sec:notations}.

\smallskip

Our general purpose oracle, with a tuneable number of ancillas, is presented in \cref{oracle:oracle_1}. We also present three special cases in \cref{oracle:oracle_2,oracle:oracle_3,oracle:oracle_4}, when the oracle further simplifies, yielding better complexities.

\smallskip

For all oracles below, let $f : \F_2^n \rightarrow \F_2^d$ and $\W_f \eqdef \left| \supp \( \WH \( f \) \) \right|$.

\begin{center}
    \fbox{\begin{minipage}{0.95\linewidth}
        \begin{center} \textbf{\cref{oracle:oracle_1}} \\ {\small (tuneable number of clean ancillas \& arbitrary Walsh--Hadamard Transform)} \end{center}
        For each $l \in \left[ 0, n \right] \cap \Z$, we construct a quantum RAM oracle, $\cO_f$, for $f$, such that, for all $x \in \F_2^n$ and $y \in \F_2^d$:
        \begin{equation}
            \cO_f \ket{x}_n \ket{y}_d \ket{0}_{d \( 2^l - 1 \)} = \ket{x}_n \ket{y + f(x)}_d \ket{0}_{d \( 2^l - 1 \)}, \tag{$\mathtt{O_1}$} \label[oracle]{oracle:oracle_1}
        \end{equation}
        The worst case complexities of \cref{oracle:oracle_1} are as follows:
        \begin{center}
            \begin{tabular}{|l|r|}
            \hline
            Total depth     & $O \( \( l + \log_2 (d) \) 2^{n - l} \)$ \\
            $R_Z$ depth     & $2^{n - l}$ \\
            Gate count      & $O \( d 2^n \)$ \\
            $R_Z$ count     & $d \W_f$ \\
            Connectivity    & $O \( l + \log_2 (d) \)$ \\
            \hline
            \end{tabular}
        \end{center}
    \end{minipage}}
\end{center}

\smallskip

\begin{center} Special cases \end{center}

Recall that $\W_f$ is the sparsity of the Walsh--Hadamard Transforms. The second oracle uses $d \W_f$ ancillas, but has depth that is only logarithmic in $\W_f$.

\smallskip

\fbox{\begin{minipage}{0.95\linewidth}
    \begin{center} \textbf{\cref{oracle:oracle_2}} \\ {\small (maximal number of clean ancillas \& sparse Walsh--Hadamard Transform)} \end{center}
    We construct a quantum RAM oracle, $\cO_f$, for $f$, such that, for all $x \in \F_2^n$ and $y \in \F_2^d$:
    \begin{equation}
        \cO_f \ket{x}_n \ket{y}_d \ket{0}_{d \( \W_f - 1 \)} = \ket{x}_n \ket{y + f(x)}_d \ket{0}_{d \( \W_f - 1 \)}, \tag{$\mathtt{O_2}$} \label[oracle]{oracle:oracle_2}
    \end{equation}
    The worst case complexities of \cref{oracle:oracle_2} are as follows:
    \begin{center}
        \begin{tabular}{|l|r|}
        \hline
        Total depth     & $O \( \log_2 \( \W_f \) + \log_2 (d) \)$ \\
        $R_Z$ depth     & $1$ \\
        Gate count      & $O \( d \W_f \)$ \\
        $R_Z$ count     & $d \W_f$ \\
        Connectivity    & $O \( \log_2 \( \W_f \) + \log_2 (d) \)$ \\
        \hline
        \end{tabular}
    \end{center}
\end{minipage}}

\smallskip

The third oracle uses no ancillas and is advantageous when $\W_f$ is much smaller than the number of nonzero values of $f$.

\smallskip

\begin{center}
    \fbox{\begin{minipage}{0.95\linewidth}
        \begin{center} \textbf{\cref{oracle:oracle_3}} \\ {\small (no ancillas \& sparse Walsh--Hadamard Transform)} \end{center}
        We construct a quantum RAM oracle, $\cO_f$, for $f$, such that, for all $x \in \F_2^n$ and $y \in \F_2^d$:
        \begin{equation}
            \cO_f \ket{x}_n \ket{y}_d = \ket{x}_n \ket{y + f(x)}_d, \tag{$\mathtt{O_3}$} \label[oracle]{oracle:oracle_3}
        \end{equation}
        The worst case complexities of \cref{oracle:oracle_3} are as follows:
        \begin{center}
            \begin{tabular}{|l|r|}
            \hline
            Total depth     & $O \( \log_2 \( \max \( \left\{ n, d \right\} \) \) \W_f \)$ \\
            $R_Z$ depth     & $\W_f$ \\
            Gate count      & $O \( d \W_f \)$ \\
            $R_Z$ count     & $d \W_f$ \\
            Connectivity    & $O \( \log_2 (d) \)$ \\
            \hline
            \end{tabular}
        \end{center}
    \end{minipage}}
\end{center}

\smallskip

The fourth and last oracle is a further specialization of \cref{oracle:oracle_3}, when the nonzero components of the Walsh--Hadamard Transform concentrate around very small and very large degrees, but avoid degrees in the middle (that is, when the degree $\approx \tfrac{n}{2}$). This method uses one ancilla when there are large degree components.

\smallskip

\begin{center}
    \fbox{\begin{minipage}{0.95\linewidth}
        \begin{center} \textbf{\cref{oracle:oracle_4}} \\ {\small (one clean ancilla \& bounded degree)} \end{center}
        Let $k \leqslant \tfrac{n}{2}$ and suppose that $\supp \( \WH \( f \) \)$ contains no bitstrings of Hamming weight strictly between $k$ and $n - k$. We construct a quantum RAM oracle, $\cO_f$, for $f$, such that, for all $x \in \F_2^n$ and $y \in \F_2^d$:
        \begin{equation}
            \cO_f \ket{x}_n \ket{y}_d \ket{0}_1 = \ket{x}_n \ket{y + f(x)}_d \ket{0}_1, \tag{$\mathtt{O_4}$} \label[oracle]{oracle:oracle_4}
        \end{equation}
        The worst case complexities of \cref{oracle:oracle_4} are as follows:
        \begin{center}
            \begin{tabular}{|l|r|}
            \hline
            Total depth     & $O \( \log_2 (d) \W_f \)$ \\
            $R_Z$ depth     & $\W_f$ \\
            Gate count      & $O \( d \W_f \)$ \\
            $R_Z$ count     & $d \W_f$ \\
            Connectivity    & $O \( \log_2 \( n \) + \log_2 (d) \)$ \\
            \hline
            \end{tabular}
        \end{center}
        {\tiny \textbf{Note:} In this case $\W_f \leqslant 2 B_{n, k} < 1.96 \cdot 2^{H_2 \( \nicefrac{k}{n} \) n}$, where $B_{n, k}$ is defined in \cref{eq:bnk}; cf. \cite{gottlieb_vc_2012}*{Lemma 7.1}.}
    \end{minipage}}
\end{center}

\bigskip

\section{Designs}
\label{sec:design}

The classical preprocessing of the function, $f: \F_2^n \rightarrow \F_2^d$, involves computing the Walsh--Hadamard Transform, $\WH \( f \) : \F_2^n \rightarrow \Z$, as in \cref{eq:WHT}. This can be done classically via the \emph{Fast Walsh--Hadamard Transform} in $O \( d n 2^n \)$ time.

In all the design patterns below we assume that the first $n + d$ qubits are in an arbitrary state, but the remaining ancilla qubits, if there are any, are initialized in the all-zero state. Miscellaneous gates are defined in \Cref{sec:pfo}.

Let $z_1, z_2, \ldots, z_{2^l}$ be (any) enumeration of $\F_2^l$. Let $0_l = u_1, u_2, \ldots, u_{2^{n - l}}$ be a periodic Gray code for $\F_2^{n - l}$ (thus, $u_{2^l + 1} = u_1$), and for each $k \in \Z$, let $\iota_{n - l} (k) \subset \left\{ 1, 2, \ldots, n - l \right\}$ be the indices of the bit changes from $u_i$ to $u_{i + 1}$.

\begin{center}
    \fbox{\begin{minipage}{0.95\linewidth}
        \begin{center} \textbf{\underline{\cref{oracle:oracle_1}}} \end{center}
        \begin{enumerate}[leftmargin=0.05\linewidth,rightmargin=0.01\linewidth]
            \item For all $j \in \left\{ 1, 2, \ldots, d \right\}$, apply an $R_Z$ gate with angle $\tfrac{\pi \( 1 - 2^d \)}{2^j}$ to the $\( n + j \)^{\mathrm{th}}$ qubit.
            \item Apply the $d$-qubit Quantum Fourier Transform to the qubits starting from $n + 1$ to $n + d$.
            \item Apply the $\A_{\F_2^l \times \{ 0_{n - l} \}, d}$ gate from \cref{eq:ASd}, controlled by the first $l$ qubits and targeting the last $d 2^l$ qubits.
            \item For all $i \in \left\{ 1, 2, \ldots, 2^l \right\}$ and $j \in \left\{ 1, 2, \ldots, d \right\}$, apply an $R_Z$ gate with angle $\tfrac{2 \pi}{2^{n + j}} \WH \( f \) (z_i, 0_{n - l})$ to the $\( n + \( i - 1 \) d + j \)^{\mathrm{th}}$ qubit.
            \item Starting from $k = 1$ to $k = 2^{n - l}$, do:
            \begin{itemize}
                \item[(A)] Apply a $\PFO_{\( u_{k - 1} \oplus u_k, d 2^l \)}$ gate controlled by the first $n$ qubits and targeting the last $d 2^l$ qubits.
                \item[(B)] For all $i \in \left\{ 1, 2, \ldots, 2^l \right\}$ and $j \in \left\{ 1, 2, \ldots, d \right\}$, apply an $R_Z$ gate with angle $\tfrac{2 \pi}{2^{n + j}} \WH \( f \) (z_i, u_k)$ (if this number is nonzero) to the $\( n + \( i - 1 \) d + j \)^{\mathrm{th}}$ qubit.
            \end{itemize}
            \item Apply a fan-out gate controlled by the $l + \iota_{n - l} \( 2^{n - l} \)^{\mathrm{th}}$ qubit and targeting the last $d 2^l$ qubits.
            \item Apply the inverse of the $\A_{\F_2^l \times \{ 0_{n - l} \}, d}$ gate controlled by the first $l$ qubits and targeting the last $d 2^l$ qubits.
            \item Apply the $d$-qubit Inverse Quantum Fourier Transform to the qubits starting from $n + 1$ to $n + d$.
            \item For all $j \in \left\{ 1, 2, \ldots, d \right\}$, apply an $R_Z$ gate with angle $- \tfrac{\pi \( 1 - 2^d \)}{2^j}$ to the $\( n + j \)^{\mathrm{th}}$ qubit.
        \end{enumerate}
    \end{minipage}}
\end{center}

\smallskip

Let $z_1, z_2, \ldots, z_{\W_f}$ be an enumeration of $\supp \( \WH \( f \) \)$.

\begin{center}
    \fbox{\begin{minipage}{0.95\linewidth}
        \begin{center} \textbf{\cref{oracle:oracle_2}} \end{center}
        \begin{enumerate}[leftmargin=0.05\linewidth,rightmargin=0.01\linewidth]
            \item For all $j \in \left\{ 1, 2, \ldots, d \right\}$, apply an $R_Z$ gate with angle $\tfrac{\pi \( 1 - 2^d \)}{2^j}$ to the $\( n + j \)^{\mathrm{th}}$ qubit.
            \item Apply the $d$-qubit Quantum Fourier Transform to the qubits starting from $n + 1$ to $n + d$.
            \item Apply the $\A_{\supp \( \WH \( f \) \), d}$ gate from \cref{eq:ASd}, controlled by the first $n$ qubits and targeting the last $d \W_f$ qubits.
            \item For all $i \in \left\{ 1, 2, \ldots, \W_f \right\}$ and $j \in \left\{ 1, 2, \ldots, d \right\}$, apply an $R_Z$ gate with angle $\tfrac{2 \pi}{2^{n + j}} \WH \( f \) (z_i)$ to the $\( n + \( i - 1 \) d + j \)^{\mathrm{th}}$ qubit.
            \item Apply the inverse of the $\A_{\supp \( \WH \( f \) \), d}$ gate controlled by the first $l$ qubits and targeting the last $d 2^l$ qubits.
            \item Apply the $d$-qubit Inverse Quantum Fourier Transform to the qubits starting from $n + 1$ to $n + d$.
            \item For all $j \in \left\{ 1, 2, \ldots, d \right\}$, apply an $R_Z$ gate with angle $- \tfrac{\pi \( 1 - 2^d \)}{2^j}$ to the $\( n + j \)^{\mathrm{th}}$ qubit.
        \end{enumerate}
    \end{minipage}}
\end{center}

\smallskip

\begin{remark}
    When $l = n$ and $\supp \( \WH \( f \) \) = \F_2^n$, \cref{oracle:oracle_1,oracle:oracle_2} are the same.
\end{remark}

\smallskip

Let $z_0 \eqdef 0$. Let $z_1, z_2, \ldots, z_{\W_f}$ be an enumeration of $\supp \( \WH \( f \) \)$ that is compatible with a Gray code.

\smallskip

\begin{center}
    \fbox{\begin{minipage}{0.95\linewidth}
        \begin{center} \textbf{\cref{oracle:oracle_3}} \end{center}
        \begin{enumerate}[leftmargin=0.05\linewidth,rightmargin=0.01\linewidth]
            \item For all $j \in \left\{ 1, 2, \ldots, d - 1 \right\}$, apply an $R_Z$ gate with angle $\tfrac{\pi \( 1 - 2^d \)}{2^j}$ to the $\( n + j \)^{\mathrm{th}}$ qubit.
            \item Apply the $d$-qubit Quantum Fourier Transform to the last $d$ qubits.
            \item Starting from $i = 1$ to $i = \W_f$, do:
                \begin{itemize}
                    \item[(A)] Apply a $\PFO_{z_{i - 1} \oplus z_i, d}$ gate as in \cref{eq:pfo}.
                    \item[(B)] For $j \in \left\{ 1, 2, \ldots, d - 1 \right\}$, apply an $R_Z$ gate with angle $\tfrac{2 \pi}{2^{n + j}} \WH \( f \) (z_i)$ to the $\( n + j \)^{\mathrm{th}}$ qubit.
                \end{itemize}
            \item Apply a $\PFO_{z_{\W_f}, d}$ gate.
            \item Apply the $d$-qubit Inverse Quantum Fourier Transform to the last $d$ qubits.
            \item For all $j \in \left\{ 1, 2, \ldots, d \right\}$, apply an $R_Z$ gate with angle $- \tfrac{\pi \( 1 - 2^d \)}{2^j}$ to the $\( n + j \)^{\mathrm{th}}$ qubit.
        \end{enumerate}
    \end{minipage}}
\end{center}

\smallskip

\begin{remark}
    When $l = 1$ and $\supp \( \WH \( f \) \) = \F_2^n$, \cref{oracle:oracle_1,oracle:oracle_3} are the same.
\end{remark}

\medskip

Recall the definition of $B_{n, k}$ from \cref{eq:bnk}. Let $k < \tfrac{n}{2}$ and let $z_1, z_2, \ldots, z_{B_{n, k}}$ be the enumeration of $n$-dimensional binary vectors with Hamming weight at most $k$, in a way that neighboring elements differ by at most two bits. Such codes exist; cf. \Cref{app:poly_gray}. Redefine $\iota_f$ accordingly. Let $\neg z \in \F_2^n$ be the bitwise negation of the bitstring $z \in \F_2^n$. Note that $h(\neg z) = n - h(z)$.

\smallskip

\begin{center}
    \fbox{\begin{minipage}{0.95\linewidth}
        \begin{center} \textbf{\cref{oracle:oracle_4}} \end{center}
        Initialize the ancilla in the zero state and use a parity-fan-out gate controlled by the first $n$ qubits and targeting the ancilla. After that repeat the design for \cref{oracle:oracle_3}, with the following two differences:
        \begin{itemize}[leftmargin=0.05\linewidth,rightmargin=0.01\linewidth]
            \item The order of components is the one given above.
            \item Wen $\WH \( f \) (\neg z_i) \neq 0$, Step $3$ has the following three further parts:
                \begin{enumerate}
                    \item[(C)] Apply a fan-out gate, controlled by the ancilla and targeting the last $d$ qubits.
                    \item[(D)] For $j \in \left\{ 1, 2, \ldots, d \right\}$, apply an $R_Z$ gate with angle $\tfrac{2 \pi}{2^{n + j}} \WH \( f \) (\neg z_i)$ to the $\( n + j \)^{\mathrm{th}}$ qubit.
                    \item[(E)] Apply a fan-out gate, controlled by the ancilla and targeting the last $d$ qubits.
                \end{enumerate}
        \end{itemize}
    \end{minipage}}
\end{center}

\smallskip

\begin{remark}
    When a projective quantum RAM oracle is needed only, that is, when
    \begin{equation}
        \cO_f \ket{x}_n \ket{y}_d = e^{i \alpha_{f, x}} \ket{x}_n \ket{y \boxplus f(x)}_d,
    \end{equation}
    is enough, for some $\alpha_{f, x} \in \rl$, then the first and the last steps can be omitted.
\end{remark}

\smallskip

\begin{remark}
    When $y = 0$ can be assumed, then Step $1$ can be omitted and the initial Quantum Fourier Transform in Step $2$ can be replaced by the simpler $H^{\otimes d}$ gates, the \emph{Quantum Walsh--Hadamard Transform}.
\end{remark}

\smallskip

\begin{remark}
    The circuit below implements \cref{oracle:oracle_3} for $n = d = 2$ and
    \begin{equation}
        f \( 0, 0 \) = 1, \quad f \( 1, 0 \) = 0, \quad f \( 0, 1 \) = - 2, \quad \& \quad f \( 1, 1 \) = 1.
    \end{equation}
    
    \includegraphics[width=\textwidth]{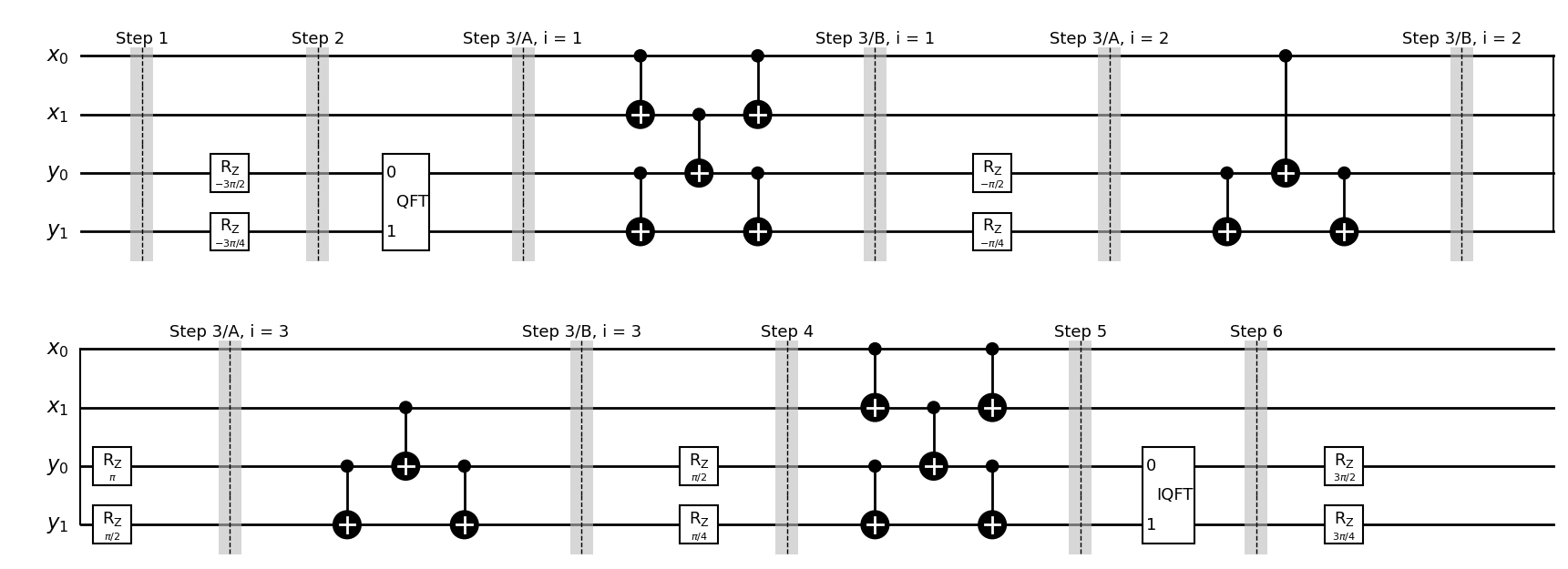}
\end{remark}

\bigskip

\section{Fault-tolerant implementation}
\label{sec:fault}

In this section we analyze the necessary overhead coming from quantum error correction when the oracles are in an approximate Clifford + $T$ decomposition.

\smallskip

First we discuss approximate Clifford + $T$ decompositions of the oracles. As out gate set is $\left\{ H, \CNOT, R_Z, \QFT_d, \QFT_d^\dagger \right\}$, the need for approximate implementations comes from the non-Clifford part, $\left\{ R_Z, \QFT_d, \QFT_d^\dagger \right\}$. The Quantum Fourier Transform and its inverse are only used once each, and they can be implemented, up to $\epsilon_\QFT$ precision, with a Clifford + $T$ count of $O \( d \log_2 \( \nicefrac{d}{\epsilon_\QFT} \) \)$; cf. \cite{nam_approximate_2020}.

\smallskip

Thus, the main contribution comes from the $R_Z$ gates. In our construction, these all have the form $R_Z \( \tfrac{\pi}{2^{n + j}} k \)$, where $k, j$ are integers and $j$ runs from $0$ to $d - 1$. Such gates can only be $\epsilon$-approximated in depth $O \( \log_2 \( \nicefrac{1}{\epsilon} \) \)$; cf \cite{neil_optimal_2016}. Fix $\epsilon > 0$ and let $\epsilon_0 \eqdef \tfrac{\epsilon}{3 d \W_f} = \tfrac{\epsilon}{3 R_Z \textnormal{ count}}$. Then approximating every $R_Z$ gate up to $\epsilon_0$ precision results in $\epsilon$-approximations of the oracles and the time complexity increase is only $O \( \log_2 \( \nicefrac{d \W_f}{\epsilon} \) \)$-fold. Thus, setting $\epsilon_\QFT = \tfrac{\epsilon}{3}$ for both the Quantum Fourier Transform and its inverse, the $T$ depths dominates the total depths (which we now use as our proxy for time-complexity) which change as:
\begin{itemize}

    \item In the case of \cref{oracle:oracle_1}, the depth increase is:
    \begin{equation}
        O \( \( l + \log_2 (d) \) 2^{n - l} \) \rightarrow O \( \( \log_2 \( \W_f \) + l + \log_2 (\nicefrac{d}{\epsilon}) \) 2^{n - l} \).
    \end{equation}
    In particular, when $n = l$, we have that the depth is $O \( n + \log_2 (\nicefrac{d}{\epsilon}) \)$.

    \item In the case of \cref{oracle:oracle_2}, the depth increase is:
    \begin{equation}
        O \( \log_2 \( \W_f \) + \log_2 (d) \) \rightarrow O \( \log_2 \( \W_f \) + \log_2 (\nicefrac{d}{\epsilon}) \).
    \end{equation}

    \item In the case of \cref{oracle:oracle_3}, the depth increases are the same:
    \begin{equation}
        O \( \log_2 \( \max \left\{ n, d \right\} \) \W_f \) \rightarrow O \( \( \log_2 \( \W_f \) + \log_2 (\nicefrac{\max \left\{ n, d \right\}}{\epsilon}) \) \W_f \).
    \end{equation}

    \item In the case of \cref{oracle:oracle_4}, the depth increases are the same:
    \begin{equation}
        O \( \log_2 (d) \W_f \) \rightarrow O \( \( \log_2 \( \W_f \) + \log_2 (\nicefrac{d}{\epsilon}) \) \W_f \).
    \end{equation}

\end{itemize}

\smallskip

Implementing a fault-tolerant protocol for our oracles would require an overhead that is logarithmic in the above complexities. Typical uses of quantum RAM oracles include, for example, quantum state preparation, Grover's algorithm, and Simon's algorithm. In all of those cases there can be (at most) exponentially many calls to the oracle. Thus, to keep the overall error bounded, one needs $\epsilon = O \( 2^{- \alpha n} \)$, for some $\alpha > 0$. For example, in the case of Grover search, one needs $\alpha > \tfrac{1}{2}$; cf. \cite{regev_impossibility_2008}. Choosing, for example, $\epsilon = 2^{- \( \frac{1}{2} + \beta \) n}$, for any $\beta > 0$, the complexities above change by replacing $\log_2 \( \W_f \) \rightarrow n$.

\medskip

Let us conclude this section with a few observations.

\begin{remark}

    As the sole source of non-Clifford gates are the $R_Z$ gates, and from this source, the dominant contribution comes from gates of the form $R_Z \( \tfrac{\pi}{2^{n + d}} k \)$, with $|k|$ being a small integer. Thus, either the elimination or the more efficient implementation  of those gates can further improve our methods. Such improvements can come, for example, in the following ways:
    \begin{itemize}
        \item Angles that are $O (\epsilon_0)$ can be now omitted, resulting in further reduction in complexity and, importantly, in $T$ gate count. This suggests that for $\epsilon$ not \emph{too small} there could be significant further savings.

        \item There is no a priori reason against using hybrid encoders, that is, combining our method with previous ones. In particular, one can easily construct functions, $f$, that decompose into $f_1 + f_2$ so that $f_1$ and $f_2$ has disjoint support, $\cO_{f_1, d}$ has favorable complexity when implanted via either of \cref{oracle:oracle_1,oracle:oracle_2,oracle:oracle_3,oracle:oracle_4}, while $\cO_{f_2, d}$ has favorable complexity when implanted via either of the techniques from \cites{low_trading_2018,seidel_automatic_2023,gilliam_grover_2021}.

        \item Techniques such as \cites{campbell_efficient_2016,choi_fault_2023} can help further reduce the complexities of small angle rotations.
    \end{itemize}    
    The above points are mostly relevant in the case of \cref{oracle:oracle_1} with $l \sim n$ and especially for \cref{oracle:oracle_2}. However, a more detailed analysis of these questions are beyond the scope of this paper.    
\end{remark}

\bigskip

\section{Boolean functions with low approximate degrees}
\label{sec:approx_degree}

Let now $f$ be a boolean function, that is $f : \F_2^n \rightarrow \F_2$. Assume now that $\widetilde{f} : \F_2 \rightarrow \rl$ is a $\nicefrac{1}{4}$-approximation of $(- 1)^f$, that is
\begin{equation}
    \forall x \in \F_2^n: \quad \left| f(x) - \widetilde{f}(x) \right| < \nicefrac{1}{4}.
\end{equation}
Now $\WH \( \widetilde{f} \)$ takes values in $\left[ - \tfrac{5}{4}, \tfrac{5}{4} \right]$. Let $d_0 \eqdef \log_2 \( 12 \W_{\widetilde{f}} \)$ and truncate the fractional part of $\WH \( \widetilde{f} \)$ after $d_0$ bits. Let us call the new function $g$. Now $g$ can be exactly represented on $d_f \eqdef 2 + d_0$ bits, and it is still a $\tfrac{1}{4} + \tfrac{1}{12} = \tfrac{1}{3}$-approximation of $f$. Most importantly, the most significant bit of $g$ is equals $f$ and $\W_g = \W_{\widetilde{f}}$. Thus, if $\W_{\widetilde{f}}$ is sparser than $\W_f$, we can apply the construction in \cref{oracle:oracle_2,oracle:oracle_3} to implement the QRAM corresponding to $f$ with lower depth and gate counts, albeit at a cost of $O \( \log_2 \( \W_{\widetilde{f}} \) \)$ ancillas.

Finding sparse approximations of boolean functions has been an active problem in (quantum) complexity theory; cf. \cite{beals_quantum_2001} or more recently \cite{aaronson_degree_2021}. When the approximate degree $\widetilde{\deg} (f) = o \( \deg \( f \) \) \ll \tfrac{n}{2}$, then $\W_{\widetilde{f}} = 2^{O \( H_2 \( \nicefrac{\widetilde{\deg} (f)}{n} \) n \)}$, implying
\begin{equation}
    d_f = O \( H_2 \( \nicefrac{\widetilde{\deg} (f)}{n} \) n \) = O \( \widetilde{\deg} (f) \log_2 \( \nicefrac{n}{\widetilde{\deg} (f)} \) \).
\end{equation}
Thus, applying our construction in \cref{oracle:oracle_4} yields a QRAM oracle for $f$ with depth and gate count being $O \( d_f 2^{H_2 \( \nicefrac{\widetilde{\deg} (f)}{n} \) n} \)$ and Clifford + $T$ depth $O \( \widetilde{\deg} (f) \log_2 \( \nicefrac{n}{\widetilde{\deg} (f)} \) 2^{H_2 \( \nicefrac{\widetilde{\deg} (f)}{n} \) n} \)$. In particular, in Aaronson et al. in \cite{aaronson_degree_2021} show that \emph{read-once formulas} possess $\tfrac{1}{3}$-approximations with degrees in $\Theta \( \sqrt{n} \)$. In these cases, $\W_{\widetilde{f}} = 2^{O \( \sqrt{n} \log_2 \( n \) \)}$ can be achieved, implying $d = O \( \sqrt{n} \log_2 \( n \)  \)$. Thus, applying our construction in \cref{oracle:oracle_4} yields a QRAM for $f$ with depth and gate count $O \( \sqrt{n} \log_2 \( n \) 2^{O \( \sqrt{n} \log_2 \( n \) \)} \)$.

Importantly, note that such usage of approximate boolean functions was made possible by the fact that our construction relied on the Walsh--Hadamard Transform of $f$ viewed as an integer-valued function and not as a binary vector-valued one. In particular, not just low degree approximations, but, more generally, approximations with sparse Walsh--Hadamard Transforms are useful for our purposes. Further investigation of these synergies between approximate boolean functions and out oracles is deferred to a future work.

\bigskip

\section{Alternative design using exact adders}
\label{sec:WH_adder}

The idea for the following was suggested to the first author by Witold Jarnicki: \cref{oracle:oracle_1,oracle:oracle_2,oracle:oracle_3,oracle:oracle_4} all use the decomposition of the quantum adder, $T_{k, d}^\QFT$ in \cref{eq:QFT_adder}. A major downside of $T_{k, d}^\QFT$ is that it uses arbitrary angle $R_Z$ gates and the Quantum Fourier transform, thus, Clifford + $T$ implementations are necessarily approximate and still costly. Since other implementations of $T_{k, d}$ can potentially be used as well. We can also make use of ``exact'' quantum adders, such as logarithmic depth carry-lookahead adders \cite{draper_logarithmic_2004} or Gidney's adder \cite{gidney_halving_2018}. We emphasize that the ideas below are just to showcase further alternatives, and not fully fleshed out. More detailed study of these ideas is the topic of a future work.

\smallskip

First, note that
\begin{equation}
    f(x) = \sum\limits_{z \in \F_2^n} \WH \( f \) (z) (- 1)^{x \odot z} = f \( 0_n \) - 2 \sum\limits_{z \in \F_2^n} \WH \( f \) (z) \( x \odot z \).
\end{equation}
Next, observe that as $f(x) \in \left[ 0, 2^d - 1 \right] \cap \Z$, we have that $\WH \( f \) (z) \in \left[ 0, 2^d - 1 \right] \cap \( \Z / 2^n \)$. Thus, $\WH \( f \) (z)$ can be represented on $n + d$ (qu)bits. Let us know start with a quantum computer with $n + \( n + d + 1 \) + 1 + \( n + d + 1 \) = 3 n + 2 d + 3$ qubits. The first $n + d$ qubits have the same role as before, storing $\ket{x}_n \ket{y}_d$ in the beginning, while the remaining qubits are ancillas initialized in the all-zero state. Qubits from index $n + d + 1$ to $n + d + n + 1$ store intermediate overflow values. Thus, we call qubits from index $n + 1$ to $2 n + d + 1$ the \emph{value qubits}. The $\( 2 n + d + 1 \)^{\mathrm{st}}$ qubit stores the values of $x \odot z$, thus we call it the \emph{parity qubit}, and the last $n + d + 1$ qubits store the double of the Walsh--Hadamard components ($d$ qubits for the integer part, $n$ qubits for the fractions), thus we call them the \emph{summand quits}. Now let $0_n = z_1, z_2, \ldots, z_{2^n}$ be a Gray code. Then we implement $\cO_f$ as follows:

\smallskip

\begin{center}
    \fbox{\begin{minipage}{0.95\linewidth}
        \begin{enumerate}[leftmargin=0.05\linewidth,rightmargin=0.01\linewidth]
            \item Digitize $f \( 0_n \)$ on the summand qubits and add it to the value qubits.
            \item For $k = 1$ to $k = 2^n$, do:
            \begin{itemize}
                \item[($2$/A)] Digitize $x \odot z_k$ on the parity qubit.
                \item[($2$/B)] Digitize $\WH \( f \) (z_k)$ on the last summand qubits.
                \item[($2$/C)] Controlled by the parity qubit, add summand qubits to the value qubits.
            \end{itemize}
            \item (Optional) Clear the ancillas.
        \end{enumerate}
    \end{minipage}}
\end{center}

\smallskip

Note that Steps $1$ and $2$/(A--B) are a collection of $\mathrm{NOT}$ and $\CNOT$ gates. The only non-Clifford step is the controlled-addition in Step $3$/C. The controlled exact adders can be implemented using $O \( n + d \)$ many $T$ gates. Further inspection of the back-to-back adder circuits and various other implementations of the Toffoli gate can give small, sharp bound on the $T$ gate bound. Thus, one get the total $T$ gate count to be $O \( \( n + d \) \W_f \)$, which differs by a factor of $O \( \tfrac{\( n + \log_2 (\nicefrac{d}{\epsilon}) \) d}{n + d} \)$ from the complexities for \cref{oracle:oracle_1,oracle:oracle_2,oracle:oracle_3,oracle:oracle_4}. Furthermore, the depth of a single addition can be made logarithmically shallow, albeit at the cost of using even more ancillas; cf. \cite{draper_logarithmic_2004}. Thus, under reasonable assumptions, it scales similarly to \cref{oracle:oracle_3}. We emphasize that also that this is for an \emph{exact adder} and not an approximate one.

This implementation offer a na{\"i}ve way for parallelization: Analogously to \cref{oracle:oracle_1} one can add $O \( (n + d) 2^l \)$ ancillas, digitize the values of the Walsh--Hadamard Transform and the qubits of the form $\ket{x \odot z}$, and add them up in a controlled fashion, in parallel. One round of this aggregation would take $O \( l C_{n + d} \)$ depth, where $C_{n + d}$ is the depth a single addition, while needing $2^{n - l}$ rounds. The complexities now are $O \( \tfrac{l C_{n + d}}{n} \)$-fold larger than those of \cref{oracle:oracle_1}. In the shallowest case of \cref{oracle:oracle_2}, the analogous implementation would be $O \( C_{n + d} \)$-fold deeper. This is the artifact of having to collect the summands ``by hand'', while \cref{oracle:oracle_1,oracle:oracle_2} the aggregation happen automatically and instantaneously at the level of phases. This however does not affect the $T$ gate counts, which are the same.

\smallskip

In conclusion, while this method has better $T$ gate counts, it comes with a few drawbacks; most importantly it uses more ancillas than (the ancilla-free) \cref{oracle:oracle_3,oracle:oracle_4}, while not providing (obvious) parallelization similar to \cref{oracle:oracle_1,oracle:oracle_2}.

\bigskip

\section{Conclusion}

We proposed novel constructions of quantum random access memory oracles, based on the Walsh--Hadamard Transform of the data, when viewed as an integer valued function (``array''). We found that while our oracle does not always yield more efficient designs, there are important use-cases in which \cref{oracle:oracle_1,oracle:oracle_2,oracle:oracle_3,oracle:oracle_4} have exponentially lower complexities, when compared to existing ones. Morally, the Walsh--Hadamard Transform can provide ``structure'' hidden in the data that techniques such as the SELECT method of \cite{low_trading_2018} might miss. Such cases typically involve data with sparse Walsh--Hadamard Transforms, which is the case, for example, in polynomial optimization problems \cites{gilliam_grover_2021,nagy_fixed_2023} and machine learning \cites{valle_deep_2018,yang_fine_2019,huh_low_2021}.

\bigskip

\appendix

\section{Design verifications}
\label{app:verifications}

\begin{proof}[Verification of the Design of \cref{oracle:oracle_1}:]
    Let us assume that the initial state is $\ket{x}_n \ket{y}_d \ket{0}_{d \( 2^l - 1 \)}$. After the first three steps, we have the state
    \begin{equation}
        \ket{x}_n \ket{y}_d \ket{0}_{d \( 2^l - 1 \)} \longrightarrow \frac{1}{\sqrt{2^d}} \sum\limits_{y^\prime \in \F_2^d} \overbrace{\exp \( \tfrac{2 \pi i}{2^d} y \( y^\prime - \tfrac{1}{2} \( 2^d - 1 \) \) \) \ket{x}_n \bigotimes\limits_{z \in \F_2^l} \bigotimes\limits_{j = 1}^d \ket{y_j^\prime \oplus \( x \odot (z, 0_{n - l}) \)}}^{\ket{\phi_{x, y^\prime}} \eqdef}.
    \end{equation} 
    By linearity, it is enough to understand what happens in the rest of the design to each $\ket{\phi_{x, y^\prime}}$. Applying the $R_Z$ gates in Step $4$ is
    \begin{equation}
        \ket{\phi_{x, y^\prime}} \longrightarrow \exp \( \tfrac{2 \pi i}{2^{n + d}} \sum\limits_{z \in \F_2^l} \WH \( f \) (z, 0_{n - l}) (- 1)^{x \odot (z, 0_{n - l})} \( y^\prime - \tfrac{1}{2} \( 2^d - 1 \) \) \) \ket{\phi_{x, y^\prime}}.
    \end{equation}

    Recall that $0_l = u_1, u_2, \ldots, u_{2^{n - l}}$ is a periodic Gray code for $\F_2^{n - l}$. For each $i \in \left\{ 1, 2, \ldots, 2^l \right\}$, let
    \begin{equation}
        \ket{\phi_{x, y^\prime, i}} \eqdef \left\{ \begin{array}{ll} \ket{\phi_{x, y^\prime}}, & \mbox{if } x \odot (0_l, u_i) = 0, \\ \ket{\phi_{x, \neg y^\prime}}, & \mbox{otherwise.} \end{array} \right.
    \end{equation}
    Now, using induction, for all $1 \leqslant k \leqslant 2^{n - l}$, after the $k^{\mathrm{th}}$ round of Step $5$, we get the state
    \begin{equation}
        \exp \( \tfrac{2 \pi i}{2^{n + d}} \sum\limits_{z \in \F_2^l} \sum\limits_{j = 0}^k \WH \( f \) (z, u_j) (- 1)^{x \odot (z, u_j)} \( y^\prime - \tfrac{1}{2} \( 2^d - 1 \) \) \) \ket{\phi_{x, y^\prime, i}},
    \end{equation}
    and hence at the end of Step $6$ (using $u_{2^{n - l} + 1} = u_1 = 0$ and \cref{eq:IWHT}) we have
    \begin{equation}
        \exp \( \tfrac{2 \pi i}{2^{n + d}} \sum\limits_{z \in \F_2^n} \WH \( f \) (z) (- 1)^{x \odot z} \( y^\prime - \tfrac{1}{2} \( 2^d - 1 \) \) \) \ket{\phi_{x, y^\prime}} = \exp \( \tfrac{2 \pi i}{2^d} f(x) \( y^\prime - \tfrac{1}{2} \( 2^d - 1 \) \) \) \ket{\phi_{x, y^\prime}}.
    \end{equation}
    The full state after Step $7$ is
    \begin{equation}
        \frac{1}{\sqrt{2^d}} \sum\limits_{y^\prime \in \F_2^d} \exp \( \tfrac{2 \pi i}{2^d} \( \( y + f(x) \) y^\prime - \tfrac{1}{2} \( 2^d - 1 \) f(x) \) \) \ket{x}_n \ket{y^\prime}_d \ket{0}_{d \( 2^l - 1 \)},
    \end{equation}
    and thus the state after Step $8$ is
    \begin{equation}
        \frac{1}{2^d} \sum\limits_{y^\prime \in \F_2^d} \sum\limits_{y^{\prime \prime} \in \F_2^d} \exp \( \tfrac{2 \pi i}{2^d} \( \( y + f(x) - y^{\prime \prime} \) y^\prime - \tfrac{1}{2} \( 2^d - 1 \) \( y + f(x) \) \) \) \ket{x}_n \ket{y^{\prime \prime}}_d \ket{0}_{d \( 2^l - 1 \)},
    \end{equation}
    which simplifies to
    \begin{equation}
        \exp \( \tfrac{2 \pi i}{2^d} \tfrac{1}{2} \( 1 - 2^d \) \( y + f(x) \) \) \ket{x}_n \ket{y + f(x)}_d \ket{0}_{d \( 2^l - 1 \)}.
    \end{equation}
    Thus, after the final step, we get
    \begin{equation}
        \ket{x}_n \ket{y + f(x)}_d \ket{0}_{d \( 2^l - 1 \)},
    \end{equation}
    This finishes the verifications. The claims about the complexities can be easily verified.
\end{proof}

\medskip

\begin{proof}[Verification of the Design of \cref{oracle:oracle_2}:]
    Same as for the Design of \cref{oracle:oracle_1} in the $l = n$ case, with the exception that one can omit the qubits corresponding to zero components of the Walsh--Hadamard Transform.
\end{proof}

\medskip

\begin{proof}[Verification of the Design of \cref{oracle:oracle_3}:]
    Same as for the Design of \cref{oracle:oracle_1} in the $l = 1$ case, with the exception that one can omit the zero components of the Walsh--Hadamard Transform and combine neighboring fan-out gates to parity-fan-out ones.
\end{proof}

\medskip

\begin{proof}[Verification of the Design of \cref{oracle:oracle_4}:]
    Same as for the Design of \cref{oracle:oracle_3}, with the note that components of the Walsh--Hadamard Transform with high degree are now implemented in Steps $2$/\{C, D, E\}.
\end{proof}

\bigskip

\section{Notations}
\label{sec:notations}

We regard elements of $\F_2 = \{ 0, 1 \}$ both as booleans and integers (in particular, reals) and use $\oplus$ to denote addition modulo $2$. For binary vectors $x, z \in \F_2^n$, let $x \odot z \in \F_2$ be the modulo $2$ reduction of the standard inner product of vectors, that is
\begin{equation}
    x \odot z \equiv \sum\limits_{a = 1}^n x_a z_a \: \mod \: 2.
\end{equation}

We use the following abuses of notation:
\begin{itemize}
    \item For any $y \in \F_2^d$, we identify $y$ with the number $\sum_{a = 1}^d y_a 2^{d - a} \in \left[ 0, 2^d - 1 \right] \cap \Z$. We extend this notation to functions as well.
    \item The symbol + below means addition modulo $2^d$, both between integers and bitstrings, that is if $y_1, y_2 \in \F_2^d$, then $y_1 + y_2 \defeq y \in \F_2^d$, is the unique bitstring so that
    \begin{equation}
        \sum\limits_{a = 1}^d y_{1, a} 2^{d - a} + \sum\limits_{a = 1}^d y_{2, a} 2^{d - a} \equiv \sum\limits_{a = 1}^d y_a 2^{d - a} \mod 2^d.
    \end{equation}
\end{itemize}

\smallskip

Using the above convention, let $\WH \( f \) : \F_2^n \rightarrow \Z$ be the (nonunitary) \emph{Walsh--Hadamard Transform} of $f$, that is
\begin{equation}
    \WH \( f \) (z) \eqdef \sum\limits_{x \in \F_2^n} (- 1)^{x \odot z} f(x), \label{eq:WHT}
\end{equation}
Note that
\begin{equation}
    f(x) = \tfrac{1}{2^n} \sum\limits_{z \in \F_2^n} (- 1)^{x \odot z} \WH \( f \) (z), \label{eq:IWHT}
\end{equation}
Now let
\begin{align}
    \supp \( \WH \( f \) \) &\eqdef \left\{ \: z \in \F_2^n \: \middle| \: \WH \( f \) (z) \neq 0 \: \right\}, \\
    \W_f                    &\eqdef \left| \supp \( \WH \( f \) \) \right|.
\end{align}

\smallskip

Let $h (z) \in \Z \cap \left[ 0, n \right]$ be the \emph{Hamming weight} of the vector $z \in \F_2^n$, that is the sum of its components.

\smallskip

For each $n, k \in \Z_+$ with $k \leqslant n$, let
\begin{equation}
    B_{n, k} \eqdef \sum\limits_{i = 0}^k \tbinom{n}{i}, \label{eq:bnk}
\end{equation}
let $H_2 (\alpha)$ be the \emph{binary entropy} of $\alpha \in (0, 1)$, that is
\begin{equation}
    H_2 (\alpha) \eqdef - \alpha \log_2 (\alpha) - (1 - \alpha) \log_2 (1 - \alpha) \in (0, 1). \label{eq:entropy}
\end{equation}

\bigskip

\section{Bounded Hamming weight ``Gray codes''}
\label{app:poly_gray}

Let $n, k \in \Z_+$, $n \geqslant k$, and
\begin{equation}
    \F_2^{n, k} \eqdef \left\{ \: x \in \F_2^n \: \middle| \: h(x) \leqslant k \: \right\}.
\end{equation}
Note that $\left| \F_2^{n, k} \right| = B_{n, k}$. As $\F_2^{n, n} = \F_2^n$, (any) Gray code traverse, without repetition, the set $\F_2^{n, n}$ in a way that the Hamming distance of any consecutive elements is minimal, that is, exactly $1$. When $k$ is strictly less than $n$, then such a code might not exist for $\F_2^{n, k}$. We show however, that there are \emph{bounded Hamming weight ``Gray codes''} that traverse, without repetition, the set $\F_2^{n, k}$ in a way that the Hamming distance of any consecutive elements is (at most) $2$.

We denote the $i^{\mathrm{th}}$ codeword of such a code by $\G^{n, k} (i)$ and assume that
\begin{equation}
    \G^{n, k} (1) = 0_n, \quad \G^{n, k} (2) = \delta_1, \quad \& \quad \G^{n, k} \( B_{n, k} \) = \delta_n
\end{equation}
where $\( \delta_i \)_j = \delta_{i, j}$. In order to construct $\G^{n, k}$, fix a periodic Gray code, $\G^n$, so that the above conditions hold for each $n \in \Z_+$. Since $\G^{n, n} \eqdef \G^n$ works, we can prove the claim using induction on $n$. Assume that $\G^{n, k}$ is given for some $n \geqslant k$. Now let
\begin{equation}
    \G^{n + 1, k} (i) = \left\{ \begin{array}{ll} \( \G^{n, k} (i),  0 \), & \mbox{if } 1 \leqslant i \leqslant B_{n, k}, \\ \( \G^{n, k - 1} \( B_{n + 1, k} - i + 1 \), 1 \), & \mbox{if } B_{n, k} + 1 \leqslant i \leqslant B_{n, k} + B_{n, k - 1} = B_{n + 1, k}. \end{array} \right.
\end{equation}
It is easy to check that $\G^{n + 1, k}$ satisfies the induction hypotheses.

\bigskip

\section{Parity-fan-out gates}
\label{sec:pfo}

For each $z \in \F_2^n$, let us define the \emph{parity-fan-out} gate on $n$ control and $d$ target qubits to act as
\begin{equation}
    \PFO_{z, d} \ket{x}_n \ket{y}_d \eqdef \ket{x}_n \bigotimes\limits_{j = 1}^d \ket{y_j \oplus (x \odot z)}. \label{eq:pfo}
\end{equation}
We call the $a^{\mathrm{th}}$ qubit a control qubit, if $z_a = 1$. Parity-fan-out gates can be implemented using $\approx 2 \( h (z) + d \)$ many $\CNOT$ gates and with a circuit depth of $\approx 2 \log_2 \( 1 + \max \( h (z), d \) \)$. When there is only one control qubit, we simply call the above a \emph{fan-out} gate. 

\smallskip

Similarly, let $S \subseteq \F_2^n$ and $\A_{S, d}$ be the following gate:
\begin{equation}
    \A_{S, d} \ket{x}_n \ket{y}_d \ket{0}_{d \( |S| - 1 \)} \eqdef \ket{x}_n \bigotimes\limits_{z \in S} \bigotimes\limits_{j = 1}^d \ket{y_j \oplus (x \odot z)}, \label{eq:ASd}
\end{equation}
The above gate can be implemented using on $O \( d |S| \)$ many $\CNOT$ gates and with a circuit depth of $O \( \log_2 \( |S| \) + \log_2 (d) \)$.

\bigskip

\section{Real-valued functions and Fej\'er states}
\label{sec:rvf}

We have so far viewed $f$ as a function taking values in $\left[ 0, 2^d - 1 \right] \cap \Z \cong \F_2^d$. Moreover, the values of $f$ (or, rather the values of its Walsh--Hadamard Transform) appeared only in the angles of $R_Z$ rotations. In fact, the first and last two steps, the constructions implement the ``generalized multi-controlled, multi-target $R_Z$'' gate of the form
\begin{equation}
    \widetilde{U}_f \ket{x}_n \ket{y}_d = \exp \( \tfrac{2 \pi i}{2^d} f(x) \( y - c_d \) \) \ket{x}_n \ket{y}_d,
\end{equation}
where $c_d \eqdef \tfrac{1}{2} \( 2^d - 1 \)$ and in this case $f$ need not be integer valued.

Thus, one can run the constructions of \cref{oracle:oracle_1,oracle:oracle_2,oracle:oracle_3,oracle:oracle_4} with any $f : \F_2^n \rightarrow \rl$ function. In this case the result is
\begin{equation}
    \cO_f \ket{x}_n \ket{y}_d = \ket{x}_n \ket{y + f(x)}_{d, F},
\end{equation}
where for each $t \in \rl / \( 2^d \Z \)$, the \emph{$t$-Fej\'er state} is defined as
\begin{equation}
    \ket{t}_{d, F} \eqdef \sum\limits_{y \in \F_2^d} \Phi_d \( t - y \) \ket{y}_d, \label{eq:fejer_state}
\end{equation}
where $\Phi_d$ is the \emph{Dirichlet kernel}:
\begin{equation}
    \Phi_d \( s \) \eqdef \tfrac{1}{2^d} \sum\limits_{z = 0}^{2^d - 1} \exp \( \tfrac{2 \pi i}{2^d} s \( z - c_d \) \) = \frac{\sinc \( \pi s \)}{\sinc \( \tfrac{\pi}{2^d} s \)}.
\end{equation}
In particular, if $s$ is a nonzero integer, then $\Phi_d \( s \) = 0$, and thus, for all $t \in \Z$, $\ket{t}_{d, F} = \ket{t}_d$. Furthermore, the probability of finding $\ket{t}_{d, F}$ in the state $\ket{y}_d$ is given by the \emph{discrete Fej\'er distribution}:
\begin{equation}
    \P_{d, F} \( y | t \) \eqdef \left| {}_d\left\langle y \middle| t \right\rangle_{d, F} \right|^2 = \left| \Phi_d \( t - y \) \right|^2 = \frac{\sinc^2 \( \pi \( t - y \) \)}{\sinc^2 \( \tfrac{\pi}{2^d} \( t - y \) \)}.
\end{equation}
Note that
\begin{equation}
    \P_{d, F} \( \left\lfloor t \right\rfloor | t \) + \P_{d, F} \( \left\lceil t \right\rceil | t \) \geqslant \frac{8}{\pi^2} > 81\%,
\end{equation}
and asymptotically we have that
\begin{equation}
   \P_f \( y^\prime \middle| x, y \) \approx \frac{4 \sin^2 \( \pi r(x) \)}{\pi^2 \( f(x) + y - y^\prime \)^2}.
\end{equation}

    \bibliography{references}

\end{document}